**Methods to stabilize aqueous supercooling identified by use of an isochoric nucleation detection (INDe) device**

Anthony Consiglio*, Drew Lilley, Ravi Prasher, Boris Rubinsky, Matthew J. Powell-Palm†*

*Department of Mechanical Engineering, University of California at Berkeley, Berkeley, CA, USA.*

†*To whom correspondence should be addressed:* mpowellp@berkeley.edu

*Equal contribution to this work*

## Abstract

Stable aqueous supercooling has shown significant potential as a technique for human tissue preservation, food cold storage, conservation biology, and beyond, but its stochastic nature has made its translation outside the laboratory difficult. In this work, we present an isochoric nucleation detection (INDe) platform for automated, high-throughput characterization of aqueous supercooling at >1 mL volumes, which enables statistically-powerful determination of the temperatures and time periods for which supercooling in a given aqueous system will remain stable. We employ the INDe to investigate the effects of thermodynamic, surface, and chemical parameters on aqueous supercooling, and demonstrate that various simple system modifications can significantly enhance supercooling stability, including isochoric (constant-volume) confinement, hydrophobic container walls, and the addition of even mild concentrations of solute. Finally, in order to enable informed design of stable supercooled biopreservation protocols, we apply a statistical model to estimate stable supercooling durations as a function of temperature and solution chemistry, producing proof-of-concept supercooling stability maps for four common cryoprotective solutes.

## Introduction

The stable equilibrium freezing point of liquid water, perhaps the most studied substance on Earth, is well known to be 0°C. However, water may continue to exist in a metastable liquid state well below this temperature, and this phenomenon, termed supercooling, plays an integral role in numerous environmental[1,2], biological[3,4], medical[5–8], agricultural[9,10], and industrial contexts[11–13]. Of particular interest, stable long-term supercooling has recently been deployed in a series of successful human organ and tissue preservation studies[5–8], providing a method of holding sensitive biologics at sufficiently low temperatures to arrest expiration whilst protecting them from lethal ice formation, which is essential to increasing the accessibility and efficacy of life-saving transplantation procedures[14–16]. However, despite the broad relevance of aqueous supercooling, it has thus far been minimally characterized at the bulk (> 1 mL) volumes relevant to many applications, and as such, design of translatable supercooling protocols and devices has proven challenging.

The central challenge posed by the use of supercooling is the stochastic nature of ice formation[17]. At its core, nucleation of a solid phase is a kinetic process driven by random molecular fluctuations within a supercooled liquid—and thus while the point after which water *can* freeze can be rigorously defined as a single temperature (0°C), the point at which water *will* freeze is a complex statistical function of the supercooling temperature, the period for which it is held at this temperature, and the water's volume or surface area[18–20].

To complicate things further, ice may nucleate from supercooled water in one of two modes: homogeneous nucleation, in which the water becomes sufficiently cold to drive spontaneous ice cluster formation in the liquid interior (occurring at approximately -40°C for pure water[21–24]), or heterogeneous nucleation in which the presence of foreign agents (particulate matter and surfaces[25–29], air bubbles[30–33], etc.) reduces the kinetic barrier to ice formation and causes nucleation to occur at significantly higher temperatures. In aqueous systems of > 1 mL volume, nucleation proceeds nigh-exclusively by the heterogeneous mode, introducing a new potential dependence of any nucleation data on the materials with which the water is interfacing during a given experiment.

Given this stochastic and context-dependent nature of ice nucleation, a rigorous description of aqueous supercooling (sufficient to enable informed design of supercooling protocols) requires very high statistical power, necessitating tens-to-hundreds of trials for each condition probed. In order to achieve these sample sizes, the majority of aqueous supercooling studies have employed microliter-and-smaller water droplets, which are monitored optically or thermally in order to detect the onset of ice nucleation/ceasing of supercooling. These studies have precisely characterized several homo- and heterogeneous nucleation processes at the microscale, but have proven challenging to scale to larger volumes due to the scale-dependent confluence of volumetric and surface effects.

Thus, in order to drive the characterization of aqueous supercooling at bulk volumes and ultimately design supercooling protocols relevant to bulk applications (such as biopreservation), supercooling studies must be performed directly on bulk-volume samples. However, this must be done without sacrificing the large sample sizes needed to secure sufficient statistical power to fully specify stochastic behaviors, and a significant technical challenge is thus presented.

In this work, we introduce the isochoric nucleation detector (INDe), an experimental platform which leverages the unique thermodynamics of isochoric systems to probe supercooling in bulk-volume aqueous media at high-throughput and high statistical power.

Over the past decade, isochoric (constant-volume) thermodynamic conditions, which are achieved by confining bulk water or solution in a rigid, high-strength container in the absence of air or any other highly compressible elements, have been demonstrated to affect the aqueous freezing process in various ways[34–36]. Most significantly, the phase equilibria that result under isochoric conditions are fundamentally different than those encountered under the conventional isobaric (constant-pressure) conditions that exist when the system is open to the atmosphere. Instead of freezing entirely at sub-zero centigrade equilibrium, as is expected under isobaric conditions, an aqueous isochoric system will freeze only partially, achieving a two-phase equilibrium configuration in which the expansion of some portion of ice drives self-pressurization of the system, depressing the freezing point of the remaining portion of the system and maintaining it in a stable liquid state. This ultimate two-phase thermodynamic destination of the system not only affects the final phase equilibria experienced, but also the kinetic nucleation and growth pathway taken to get there, and prior theoretical and experimental work has suggested that isochoric conditions may enhance the stability and reduce the variability of aqueous supercooling[35,37,38], thereby enabling not only potentially robust biopreservation and other practical applications, but reliable supercooling characterization at bulk volumes.

Herein we will detail the electro-mechanical design of the INDe device and its thermodynamic operating principles and then employ it to conduct three studies investigating multiple factors that affect supercooling in aqueous systems, including thermodynamic boundary condition, surface coating and solution chemistry. Among several key findings, we demonstrate that isochoric conditions can indeed significantly enhance the depth and stability of aqueous supercooling relative to conventional isobaric conditions; that applying a hydrophobic coating to all surfaces in contact with the bulk liquid sample can further enhance aqueous supercooling regardless of thermodynamic condition, and that various common solutes will depress the maximum degree of supercooling possible by at least as much as their according freezing point depression. Finally, we deploy a Poisson-statistics model of nucleation to calculate the induction time of nucleation (i.e., the period that the supercooled liquid will remain stable) as a function of temperature for each solution using only our maximal supercooling data as an input, providing an essential tool for the informed design of supercooled biopreservation protocols. In total, this work demonstrates both the multifaceted utility of the INDe platform for nucleation analyses and several novel means of enhancing supercooling in aqueous systems.

## Results and Discussion

### Design of an Isochoric Nucleation Detector (INDe)

In order to study bulk-volume aqueous supercooling at high-throughput, we have designed a device that leverages the unique thermodynamic behaviors of aqueous systems confined under isochoric (constant-volume) to detect nucleation at low-latency without the need for scale-variant thermal or optical detection: the isochoric nucleation detector (INDe).

At the heart of the INDe, depicted in Figure 1, is a two-part isochoric (constant-volume) chamber constructed from Aluminum-7075, chosen for its preferable combination of high strength and high thermal conductivity. The chamber has an internal volume of 5mL and an inner diameter of 0.5". A threaded plug with a tapered end forms a tight metal-on-metal seal with the corresponding mating surface on the chamber body, providing a sealed interior capable of withstanding pressures in excess of 200 MPa. This design feature further creates a continuous and homogenous interior surface that minimizes the potential for active nucleation sites and thus maximizes supercooling to the greatest extent possible. Flat exterior faces of the chamber allow it to be clamped between a pair of temperature control assemblies, each comprised of a two-stage thermoelectric module and standard fan-cooled CPU heat sink. To aid in temperature control and uniformity, the chamber is further surrounded by an insulation shell of 3D printed PLA plastic filled with expanding polyurethane insulating foam. Figure SI shows several assembled INDe devices.

**Pressure-based nucleation detection**

In traditional nucleation experiments, the nucleation events are often detected optically, by sensing the change in sample translucence[28,39], or thermally, by detecting the release of latent heat[40,41]. In metallic isochoric chambers, optical detection is not possible. Detection of the latent heat release is possible; however, in systems of milliliter scale or larger, the propagation of thermal energy from the nucleation site to the temperature sensor requires appreciable time and may thus lead to measurement uncertainty. In aqueous systems under isochoric conditions however, a third signature of ice nucleation exists: pressure. Due to the difference in density between ice and water, when ice begins to crystallize from supercooled aqueous media in a confined environment, its expansion generates significant hydrostatic pressure (up to approximately 210 MPa at -22°C)[34]. Thus, in an isochoric chamber, the detection of a pressure rise serves as an alternative method for nucleation detection. Furthermore, this pressure signal propagates through the sample at the speed of sound (approximately 1500 m/s in pure water), providing extremely low detection latency and affording easy scalability to increasingly large systems.

In order to detect the pressure signature of ice nucleation, a traditional pressure transducer may be employed; however, this can introduce undesirable material interfaces, undesirable complexity and expense, as well as potential compressibility issues, which may corrupt the desired isochoric conditions. Instead, the INDe, which is specifically designed to maximize supercooling stability, utilizes a high-sensitivity full-bridge strain gauge affixed to one of the exterior faces of the chamber (Figure 1b). Elevated pressure within the sealed isochoric environment causes the chamber to mildly elastically deform, which produces a clear spike in strain gauge signal. The equivalence between direct detection of pressure and detection of strain was verified by simultaneously monitoring pressure and strain during a nucleation event. As shown in Figure 2b, the two signals are nearly coincident, with sub-one second latency. This simple strain detection method has proven to be highly sensitive, and because it is independent of chamber geometry, may be readily employed in isochoric systems of varying size.

On the face of the chamber opposite the strain gauge, a hole in which a type-K thermocouple is embedded, allows for measurement of the internal temperature (+/- 0.2°C). Agreement between this embedded thermocouple and the interior liquid temperature was verified in preliminary testing.

**Transient supercooling experiments in the INDe**

Experiments to characterize supercooling are generally conducted in one of two modes: isothermal or transient. In the isothermal mode, the sample is quenched to and held at a single sub-freezing temperature, and the time required for ice to nucleate is recorded (this "induction time" is a fundamental characteristic of supercooling, and will be discussed in further detail in the coming sections)[9,39,42–44]. In the transient mode, the sample temperature is cooled continuously at a constant rate and the temperature at which nucleation occurs is recorded[28,45–47]. This value, herein referred to as the nucleation temperature, represents the limit of stability at which the induction time approaches zero, or the maximal degree of supercooling possible.

From a theoretical standpoint, the isothermal method may be preferable, as it enables direct determination of the nucleation rate of the system, $J\left[\frac{\#\ of\ nuclei}{unit\ size * unit\ time}\right]$, which represents the most fundamental parameter employed in classical nucleation theory (CNT). However, this nucleation rate can vary many orders of magnitude with small changes in temperature[18], and the induction time may thus vary from the order of seconds to the order of years with only a few degrees change in temperature, posing a significant difficulty for laboratory experimentation (especially if trials are to be repeated tens or hundreds of times in order to establish sufficient statistical power). Thus, the transient method is much more practical for high-throughput experimentation, and while the INDe can be operated in both configurations, we conduct all experiments herein in the transient mode.

Transient supercooling experiments in the INDe begin by filling and sealing the isochoric chamber. Special attention is paid to excluding any air from the chamber during assembly, as the presence of any bulk gas phase can corrupt the desired isochoric conditions by introducing increased compressibility[48], and because the gas-liquid interface may act as a potent nucleation site[30]. To form a reliable seal capable of withstanding elevated pressures, which may exceed 200 MPa if ice growth is allowed to proceed indefinitely, the plug is tightened to a torque of 45 ft-lbs. After loading, the sealed chamber is inserted into the insulation shell and secured between the temperature control assemblies.

Utilizing a custom-developed Python control dashboard, the temperature is decreased at a rate of 2±0.5°C/min via PID control of the thermoelectric modules. Cooling is continued until the onset of nucleation, which is indicated by a spike in the strain and autonomously sensed by the control software. The thermoelectric elements are then switched into heating mode and the temperature of the chamber is quickly brought back above 0°C and held for a specified time (here 5 minutes), after which the same plunge in temperature is repeated. This heating step arrests ice growth and enables the supercooling to be "reset" after each nucleation event, enabling continuous unmonitored cycling for tens or hundreds independent nucleation events over the course of several hours.

Depicted in Figure 2a are example raw temperature and strain data for several cooling and warming cycles. The base of the strain spike corresponds to the nucleation of ice from the supercooled liquid and establishes the time at which the nucleation temperature is determined. Figure 2c depicts the evolution of nucleation temperatures for a single experiment across 100 cycles, and Figure 2d shows a violin plot depiction of this same nucleation data capturing the stochastic distribution, median value, and range. Figure 2e provides the survivor curve or cumulative distribution function of this data, which represents the fraction of unfrozen samples (or runs) for a given temperature. Each of these representations offers different insight into the statistical realities of nucleation in the target substance, with the survivor curve providing the ultimate limits of the observed nucleation probabilities. Further details on all performed statistical analyses are provided in the Methods section.

**Using the INDe to probe various factors affecting aqueous supercooling**

Supercooling is a complex phenomenon affected by myriad factors, including thermodynamic boundary conditions, surface conditions, and system chemistry. The INDe provides a versatile platform for probing all of these aspects both independently and in concert, and in order to demonstrate the breadth of studies possible, we present three studies on differing factors affecting aqueous supercooling, which culminate in the presentation of useful tools for the design of effective supercooled biopreservation protocols.

**Effects of thermodynamic conditions and surface conditions on supercooling of pure water**

Recent studies have suggested that isochoric conditions may enhance the supercoolability of aqueous solutions relative to conventional isobaric (constant pressure) conditions at atmospheric pressure. Powell-Palm et al.[37] demonstrated that supercooled isochoric systems exhibit enhanced stability against macroscopic agitations including vibration, ultrasonication, drop impact and thermal cycling, albeit at only a single mild supercooling temperature (-3°C). Further studies have also suggested that isochoric confinement may increase supercoolability by increasing the energetic barrier to nucleation and suppressing other kinetic nucleation mechanisms such as cavitation[35,38].

In order to further probe the potential effects of isochoric confinement on supercooling, the INDe is employed here to characterize the supercooling limit of deionized water in three potential thermodynamic configurations: under isobaric conditions (in which the liquid is freely exposed to the atmosphere), under isobaric + oil-sealed conditions (in which the liquid is exposed to the pressure reservoir provided by the atmosphere, but is denied contact with air by an immiscible layer of oil), and under isochoric conditions (in which the liquid is denied access to the atmosphere entirely and is rigidly confined at constant volume). Isobaric experiments were performed in the same INDe chambers, yet with the plug removed, and in the oil-sealing trials, a layer of mineral oil was placed atop the water volume, as depicted in Figure 3a. Although ice nucleation in an isobaric system is not required to be accompanied by an increase in hydrostatic pressure, the strain gauges nonetheless produce a small yet distinct spike upon nucleation, likely due to rapid ice expansion in the narrow cavity. This signal proved sufficient for nucleation detection and was additionally supplemented by monitoring of the temperature rise due to the release of latent heat (which occurred within a second of the strain rise). Since the entire volume of water freezes upon nucleation in an isobaric system, in contrast to isochoric systems, the latent heat release is significant.

In addition to these varying thermodynamic conditions, we also probe the effect of two different surface conditions on pure water supercooling. As nucleation occurs heterogeneously in most real aqueous systems[17,29,49], countless studies have probed the effects of surface conditions on ice nucleation processes. In classical nucleation theory, the contact angle of the liquid phase on the containing surface captures the propensity of that surface to aid in heterogeneous nucleation, with lower contact angles (or hydrophilic surfaces) increasing the likelihood of nucleation and higher contact angles (or hydrophobic surfaces) decreasing the likelihood of nucleation[19]. Surprisingly however, while myriad studies have examined the supercooling of droplets on hydrophobic surfaces, to our knowledge no previous studies have probed the effect of fully containing $> 1$ mL volumes of water within hydrophobic walls. Thus, for each of the thermodynamic configurations mentioned above, we also probe the effect of coating the entire interior surface of the chambers with a thin layer of petrolatum, thus exposing it to exclusively hydrophobic surfaces.

For each condition, a minimum of 50 trials per chamber in three different chambers are performed, providing a minimum $n = 150$ data points. The results of these pure water experiments with both bare aluminum surfaces and hydrophobic petrolatum-coated surfaces are shown in Figure 3, and several conclusions can be made.

Firstly, for both surface conditions, bare and coated, the isochoric systems exhibit significantly lower mean nucleation temperatures than both the isobaric and isobaric oil-sealed systems. This finding supports previous theoretical suggestions that isochoric confinement increases the nucleation barrier and decreases the probability of nucleation at a given temperature[35], and is furthermore consistent with previous experimental findings that found isochoric supercooling to be more stable than its isobaric alternatives at a given sub-zero centigrade temperature[37]. Interestingly, oil-sealing produces no statistically significant effect on the observed nucleation temperature as compared to the unsealed isobaric system, seemingly contrary to previous findings[5,30]. Based on the fact nucleation occurs heterogeneously (i.e., on surfaces) in aqueous systems of this size, we may attribute this result to the small relative surface area of the water-air interface in our system, which accounts for only approximately 6% of the total enclosed surface area. Following this logic, in systems of smaller height-to-diameter aspect ratios (such as those employed in previous studies[5,30], oil-sealing may be predicted to have a more marked effect.

A further explanation for these observations may lie in the three-phase contact line (i.e., the air-water-surface and air-oil-surface interface). This interface, which is present in both isobaric and isobaric oil-sealed systems, is not present in the isochoric system due inherently to the total confinement within the aluminum chamber. Recent studies have probed nucleation kinetics at three phase contact lines and have found increased nucleation propensity at these interfaces[50–52].

Secondly, for all three thermodynamic configurations, the addition of a hydrophobic surface coating to the bare metallic walls is found to significantly depress the mean nucleation temperature. It should be noted that the amount by which the nucleation temperature decreased is very similar between the two isobaric conditions and greater under isochoric conditions, consistent with the aforementioned surface area arguments. These results suggest that systems designed for enhanced supercooling should incorporate hydrophobic coatings not only at the air-water interface, but on every surface in contact with the liquid. However, while the supercooling enhancement effect of one token hydrophobic surface coating (petrolatum) is demonstrated here, whether these effects are specific to petrolatum or to hydrophobic coatings as a whole cannot be concluded, and future work should probe the effects of various hydrophobic coatings.

Finally, among the three thermodynamic configurations, the conventional isobaric trials produce the largest standard deviations, while the standard deviations for the isobaric oil-sealed and isochoric trials are of comparable and lesser magnitude. This suggests that an exposed air interface, which is open to convection, the introduction of microscopic particulates, small local variance in pressure, etc., may introduce the potential for inconsistent nucleation sites, a result that is consistent with previous supercooling experiments[30,37], and that isochoric or oil-sealed conditions should be employed for fundamental nucleation characterization where possible.

**Effects of common cryoprotective solutes on aqueous supercooling**

While the supercooling behaviors of pure water are of fundamental interest to materials science, myriad biological, geochemical, and atmospheric supercooling processes of interest involve the incorporation of various solutes. The equilibrium freezing point depression accompanied by the addition of various solutes to water is well documented; however, the effect of these same solutes on complex kinetic processes such as supercooling is less well understood. Amongst the many studies that have probed the metastability of aqueous solutions[29,49,53–55], it is often hypothesized that the presence of solutes disrupts the hydrogen bonding network of water molecules and their ability to produce crystalline-like order, and that this disruption results in a decreasing homogeneous nucleation temperature relative to pure water[23].

To demonstrate the utility of the INDe for characterizing the effect of solutes on supercooling in bulk aqueous solutions, we perform transient supercooling experiments on binary solutions of four common cryoprotective compounds: dimethyl sulfoxide (DMSO), ethylene glycol, glycerol and propylene glycol. In order to probe the maximal possible supercooling, per the results in the previous section, these experiments are conducted under isochoric conditions in chambers coated with petrolatum. Figure 4 shows the nucleation temperature data for trials conducted at concentrations of 1 mol%, 2.5 mol% and 5 mol% of each solute. Figure 4a-d show the distributions of the experimental nucleation data in violin plot form, Figure 4e-h show survivor curves for this data, and Figure 4i-l show the weighted mean nucleation temperatures as a function of concentration. As in the preceding pure water experiments, a minimum of 50 trials/chamber in three different chambers are performed for each condition, providing a minimum n = 150 data points, which Figure 4 displays in aggregate for each solute and concentration. The chamber-by-chamber raw data for each condition (totaling 36 trials across four solutions and three concentrations) are provided in Figure S2 of the SI.

Several significant conclusions can be drawn from the data in Figure 4. Firstly, the expected trend of decreasing nucleation temperature with increasing solution concentration is observed over all tested solutions and concentrations, indicating that no unanticipated surface-solute interactions or entropic effects develop with increasing solute presence. The absolute degrees of supercooling achieved by each solution are largely similar (within an approximately 2°C span for each mol%), with 5 mol% propylene glycol providing the deepest observed supercooling at -21.6°C.

Furthermore, the nucleation temperatures observed across solutions at the relatively mild concentrations probed here suggest strong untapped potential for supercooling in the context of biopreservation, in which the duration of preservation possible is a strong function of the degree of coldness achieved. Supercooled biopreservation studies have thus far been conducted at temperatures in the -3°C to -8°C range[6–8,56]; however, our data suggest that much colder temperatures could potentially be achieved. For example, at 5 mol% (15.4 – 21.2 mass% or 2.4 – 2.5M depending upon the solute), all four solutions exhibit maximal supercooling at temperatures less than -20°C, and at 1 mol% (3.4 – 4.9 mass% depending upon the solute or approximately 0.5M) they exhibit maximal supercooling at temperatures less than -15.9 °C. Of course, safe and high-stability supercooling cannot be performed at the limit of supercooling and additional analyses are required to estimate the temperatures at which high stability is guaranteed (discussed in the following section), but the magnitudes of the nucleation temperatures shown in Figure 4 suggest that significantly colder supercooled biopreservation is possible without the incorporation of high-toxicity concentrations cryoprotectant chemicals.

In addition to solution-by-solution analysis of absolute supercooling, useful insight can be attained through comparison of the relative supercooling (i.e., the distance in temperature past the equilibrium melting point to which the solution is supercooled) between solutions. One oft-used parameterization framework, termed the lambda method[29,49,57,58], characterizes the effect of solutes on aqueous supercooling using the relation:

$$\Delta T_{nuc} = \lambda \Delta T_{melt} \tag{1}$$

wherein $\Delta T_{melt}$ is the equilibrium melting point depression of the aqueous solution, $\Delta T_{nuc} = T_{nuc,solution} - T_{nuc,water}$, is the depression of the nucleation temperature of the solution relative to the nucleation temperature of pure water, and $\lambda$ is a constant that depends on the nature of the solute as well as experimental conditions such as the presence of specific ice nucleators and sample volume[57]. Following this approach, for a set of identical experimental conditions, the relative supercooling ability of different solutions may be compared on the basis of their $\lambda$ value. It has been found that for homogeneous nucleation in microscale systems (sub-mL volumes, typically probed using droplets), $\lambda$ is approximately equal to 2[29]. Exceptions to this exist for large molecules, such as polymers and long-chain carbohydrates, which exhibit significantly non-ideal behavior in solution and have been shown to produce $\lambda$ values up to and greater than 4[29,54]. The $\lambda$ value for homogeneous nucleation at the microscale represents the upper bound for a given solution however, and for heterogeneous nucleation, $\lambda$ will decrease with increasing nucleation propensity.

Applying the lambda method, we proceed to compare the relative supercooling between solutions. Figure 5a provides both the equilibrium melting temperature and nucleation temperature curves as a function of concentration for the four studied solutions, and comparison via the lambda method is achieved in Figure 5b by plotting the degree of additional supercooling compared to pure water, $\Delta T_{nuc}$, against the equilibrium melting point depression, $\Delta T_{melt}$. Two conclusions may be drawn from this comparison.

Firstly, the ratio $\lambda = \Delta T_{nuc}/\Delta T_{melt}$ falls between 0.95 and 1.2 across all solutions tested herein. The majority of prior nucleation research evaluating solution supercooling by the lambda method has examined microscale systems, and often in the homogeneous nucleation regime. To our knowledge, this data provides the first high-statistical power lambda values for these solutions at $> 1$ mL volumes and under consistent surface conditions, and we thus suggest that an approximate value of $\lambda = 1$ may provide a sound benchmark for the scaling of bulk supercooled solutions with their melting point depression.

Secondly, Figure 5b demonstrates that while ethylene glycol, propylene glycol, and DMSO all exhibit very similar behavior, glycerol provides appreciably less supercooling per unit melting point depression. It is difficult to speculate as to the origin of this difference, however previous work in the homogeneous regime has similarly reported that glycerol provides less supercooling as compared to ethylene glycol[54].

**Calculating nucleation induction times using INDe data**

This work has thus far presented results on the supercooling of aqueous media with the interest of measuring the absolute and relative degrees of supercooling afforded to water by the addition of various solutes or the application of different thermodynamic conditions. We now turn to the oft-overlooked next step in supercooling analysis: adaptation of this material data to a useful application.

In order to facilitate supercooled biopreservation, stability of supercooling must be ensured for extended periods, typically on the order of days or weeks—which of course precludes the use of the maximal supercooling temperatures reported in the preceding section, at which ice nucleation is induced over a period of seconds. Thus, to design an effective biopreservation protocol, one must know not simply the maximum degree of supercooling possible, but the maximum degree of supercooling *at which* the induction time of nucleation (the period that the solution will remain stably supercooled preceding the emergence of the first nucleus) exceeds the desired preservation duration. While direct experimental probing of the requisite relationship between temperature and induction time can prove incredibly time-intensive, this relationship can be reliably estimated using only the survivor curve data that we have already generated herein.

Nucleation of a solid phase from a liquid phase is often analyzed through the lens of classical nucleation theory, a semi-physics-based phenomenological framework developed in the mid-20th century. However, nucleation may also be modelled as a Poisson process, an approach that has enabled significant recent progress in untangling the phase transformation kinetics of metals, ceramic materials, and phase-change energy storage materials[40,59–61].

In particular, Lilley et al.[40] have recently developed a model by which to calculate the induction time as a function of temperature for a given system using only high-throughput bench-scale nucleation data, such as that presented in Figure 4, and we here adapt this approach to estimate the induction time-temperature relationships for our solutions of interest. Within this model, the Poisson rate parameter is taken to equal to the nucleation rate, $J\left[\frac{nuclei}{s}\right]$, which may be fitted to a function of the form

$$J(T) = \gamma \Delta T^n \tag{2}$$

wherein $\Delta T$ is the degree of supercooling (i.e., $T_{melt} - T$) and the terms $\gamma$ and $n$ are empirical fitting parameters. Furthermore, the nucleation rate is related to the survivor function (the fraction of non-nucleated samples at a given temperature, as shown in Figure 4b), by

$$\chi(T) = e^{-\frac{1}{\beta}\int_{T_m}^{T} J(T)dT} = e^{-\frac{\gamma(T_m - T)^{n+1}}{\beta(n+1)}} \tag{3}$$

wherein $\beta$ is the cooling rate in $\frac{°C}{min}$. In our INDe system, the cooling rate is prescribed, and thus Equation 3 can be fitted to the experimentally measured survivor curves (an example fit is shown in Figure S3) in order to determine $\gamma$ and $n$ and obtain the nucleation rate as a function of temperature. Finally, the average induction time $\tau$ can then be calculated as $\tau = J^{-1}$.

**Mapping stability for supercooled biopreservation**

Following the aforementioned procedure, the induction time-temperature relationship is computed for the pure water trials with petrolatum coating (Figure 6a) and for the four solutes presented thus far (Figure 6b-e). The induction times are computed for each individual trial and the shaded regions provide the range of induction times for the three individual trials of each condition. The solid lines give the average of the three individual induction times. It should be noted however that the induction times are computed from experimentally-obtained survivor functions, whose empirical functional form is non-linear. Thus, this average does not capture the full predictive power of the empirical data that the shaded range does, and may vary from the true average, the computation of which would require a method of averaging a set of multi-parameter empirical cumulative distribution functions. Future work should investigate more sophisticated mathematical approaches by which to extend averaging of raw nucleation data to computed induction times or other secondary parameters.

Figure 6a offers further insight into the effect of thermodynamic conditions. Isochoric conditions, which produce the lowest nucleation temperature of the three conditions probed in Figure 3, also produce the most stable supercooling, as can be seen by comparing the predicted induction times for the three conditions at any given temperature. Interestingly, the oil-sealing, which produced no significant effect on mean nucleation temperature when compared with conventional isobaric, shows distinctly longer inductions times on average. This result may serve to support previous oil sealing experiments conducted by Huang et al.[5,30] which were conducted over extended durations and found improved long-term stability with surface sealing.

Figure 6b-e demonstrate the effects of solutes on induction time and highlight the extreme sensitivity of induction time to temperature, with shifts of only a few degrees yielding orders-of-magnitude changes in the induction time. For example, a 2.5 mol% solution of ethylene glycol, which is expected to remain supercooled for an hour at -15°C, has a predicted induction time longer than one year at -12.5°C. Similarly, at a given temperature, a slight increase in solution concentration also increases the predicted induction times by orders of magnitude.

This sensitivity highlights the difficulty of designing supercooled biopreservation protocols without rigorous advance characterization of the desired preservation solution and suggests that Figure 6a-e may be referenced directly by the interested cryobiologist for the informed design of supercooled biopreservation protocols. For example, this analysis predicts that preservation on the order of months in a solution of 5 mol% (15.4 mass%,

~2.5M) ethylene glycol may be conducted at temperatures as cold as -16°C, or in a solution of 1 mol% (4.2 mass%, ~0.5M) solution of DMSO at temperatures as low as -12°C.

In Figures 6e-h, we extend further the utility of this average induction time data by incorporating a continuous concentration axis, which we achieve by fitting a three-dimensional surface to the curves shown in Figure 6a-e and constructing concentration-temperature-induction time heatmaps. We term these "supercooling stability maps", and the discrete contours shown capture the estimated temperature-variance of a given induction time (1 day, 1 week, 1 month, etc.) with solution concentration, providing a new supercooled biopreservation design tool. The white dashed lines represent the equilibrium melting temperature, above which a solution is indefinitely stable. As more granular understanding of the principle biological factors affecting biopreservation (solution toxicity, temperature dependence of metabolism, etc.) emerges, it is anticipated further that this data may be used to optimize preservation temperature and preservation period / induction time against solution toxicity, which is in many cases a clear function of concentration.

Some limitations to the interpretation of this data should be noted: All supercooling tests herein probed a 5 mL volume of aqueous media, and the induction times presented in Figure 6 describe this particular system. If the assumption is made that nucleation is initiated on the interior chamber surface, the computed nucleation rates and induction times may be linearly scaled by the surface area for any isochoric system with petrolatum-coated surfaces. However, future studies must experimentally validate this scaling and investigate the possibility of a volumetric contribution.

Furthermore, as noted previously, determination of the true average induction times across samples/systems, the nucleation behaviors of which all produce Poisson distributions but not the *same* Poisson distribution, proves mathematically non-trivial, and must be investigated further. What's more, the end application of supercooled biopreservation provokes many questions about the interpretation of the temperature-induction time relationship in protocol design— *should* an average induction time be used to design a biopreservation protocol? Can factors of safety be incorporated in supercooling protocol design by operating some distance from these averages, or outside the range of observed values? How can we further quantify the relative stability or instability of supercooling for a given time period as we increase or decrease temperature? These and other questions arising from the development of the proof-of-concept stability maps shown herein point to the need for significant future statistical analyses by which certainty and safety in stable supercooling can be more specifically guaranteed.

## Conclusions

In this work, a new device for high-throughput characterization of aqueous supercooling in bulk-volumes is presented, termed the isochoric nucleation detector (INDe). This device uses a new non-invasive pressure-based nucleation detection mechanism enabled by the unique thermodynamics of aqueous isochoric systems and provides a platform for probing many of the myriad factors that affect bulk-volume aqueous supercooling with high statistical power. Over the course of three studies, totaling thousands of nucleation detections, we identify three key factors that affect the stability and depth of supercooling in bulk aqueous systems. Firstly, isochoric thermodynamic conditions enable significantly deeper supercooling than conventional isobaric or isobaric oil-sealed conditions; secondly, applying a hydrophobic coating (here petrolatum) to all solid surfaces in contact with the liquid enhances supercooling regardless of thermodynamic condition; and thirdly, common cryoprotective solutes enhance the maximal supercooling temperature possible at a rate roughly equal to their freezing point depression. In order to increase the direct utility of these findings to the biopreservation community and others seeking to harness stable aqueous supercooling, we also input our maximal supercooling data into a Poisson statistics model that enables prediction of the relationship between supercooling temperature and nucleation induction time, or how long a supercooled system will remain stable at a given temperature. Finally, we use this information to introduce and construct proof-of-concept supercooling stability maps, a new reference tool to enable informed design of stable supercooled biopreservation protocols. This work in sum presents a new experimental and theoretical pipeline by which to first characterize and ultimately utilize aqueous supercooling at $>$ 1 mL volumes.

## Materials and Methods

**Experimental materials.** Deionized water (type II, SKU S25293) and ethylene glycol (SKU E178) were purchased from Thermo Fisher Scientific (USA). Mineral oil for the oil-sealing experiments (SKU M8410), propylene glycol (1,2-propanediol, SKU 398039), DMSO (dimethyl sulfoxide, SKU D5879) and glycerol (SKU G7893) were purchased from Sigma-Aldrich (USA). Petrolatum used to coat the interior chamber surface was purchased from Vaseline, Unilever (UK).

**Isochoric nucleation detector electronics.** The isochoric nucleation detection (INDe) system is comprised of temperature control assemblies and temperature and strain monitoring systems, which are controlled via a Python-based control software running on a Raspberry Pi 4B single board computer (Raspberry Pi Foundation, UK). The temperature control assemblies are each comprised of a two-stage thermoelectric module (CUI Devices CP60H-2 Series) and fan-cooled CPU heat sink (Cooler Master). The thermoelectric modules are controlled by a PID temperature controller (Opt Lasers, TEC-8A-24V-PID-HC-RS232). Full bridge aluminum strain gauges (3147_0), PhidgetBridge strain gauge DAQ (1046_0B), Thermocouple Phidget DAQ (TMP1101_0), and USB VINT Hub (HUB0000_0) were purchased from Phidgets Inc. (CA).

**Isochoric chamber loading procedure.** Solutions are first prepared using an analytical balance (A&D ER-182A). The solution (or deionized water) is then dispensed slowly into the chamber using a syringe so to not introduce any air bubbles or air pockets. The plug is then threaded into the chamber until the sealing surfaces contact each other, after which a digital torque wrench (Yellow Jacket 60648) is used to apply a torque of 45 ft-lbs. This torque is applied to ensure a tight metal-on-metal seal is formed. No pressure is applied to the liquid during this process, as excess liquid is forced out through the weep hole. This was confirmed in preliminary trials using a pressure transducer, as shown in Fig. 2c.

**Chamber surface coating.** For the coated deionized water experiments and all aqueous solution experiments, the interior chamber surface is coated with a thin layer of petrolatum. To apply this coating, the chamber is first heated using a standard heat gun, and a small amount of petrolatum (<1mL) is then placed into the chamber. After melting the petrolatum, the chamber is then inverted and simultaneously rotated in order to coat the entire surface while allowing excess liquid to drain out. The chamber is then left to cool in a refrigerator at ~2°C to allow the petrolatum coating to solidify. A thin layer of petrolatum is also applied to the bottom surface of the plug.

**Statistical analysis.** For all supercooling results presented, a minimum of 50 consecutive nucleation cycles were performed for each trial and repeated in three separate chambers, totaling a minimum of $n=150$ data points per experimental condition. These data are then aggregated for each condition and reported as a weighted average across trials, accompanied by a weighted average of the standard deviations of each trial. Statistical difference was confirmed using one-way ANOVA and Bonferroni's multiple comparisons test. Significant difference was indicated by a value of $p < 0.05$, and groups that were determined not to be significantly different (i.e., share a common mean) were marked with asterisks. Due to the large number of data points obtained for each condition, the standard error of the mean was miniscule in all cases and was thus not reported.

## Acknowledgements

This work received financial support from the National Science Foundation (NSF) Graduate Research Fellowship under Grant No. DGE 1752814 as well by the NSF Engineering Research Center for Advanced Technologies for Preservation of Biological Systems (ATP-Bio) under NSF EEC Grant No. 1941543.

## Author Contributions

A.C., B.R., and M.P.P. conceived of and designed the study. A.C. and M.P.P. designed and built the devices, performed the experiments, and analyzed and interpreted the data. D.L. developed and applied the statistical model, with input from R.P. A.C. and M.P.P. wrote the manuscript. B.R. and R.P. provided critical revisions to the manuscript. All authors participated in interpretation of the results and editing of the manuscript.

## Competing Interest

M.J.P.P. and B.R. have a financial stake in BioChoric Inc. and both they and the company may benefit from commercialization of the results of this research. The remaining authors declare no conflict of interest.

## Data Availability

All data are available from the authors upon reasonable request.

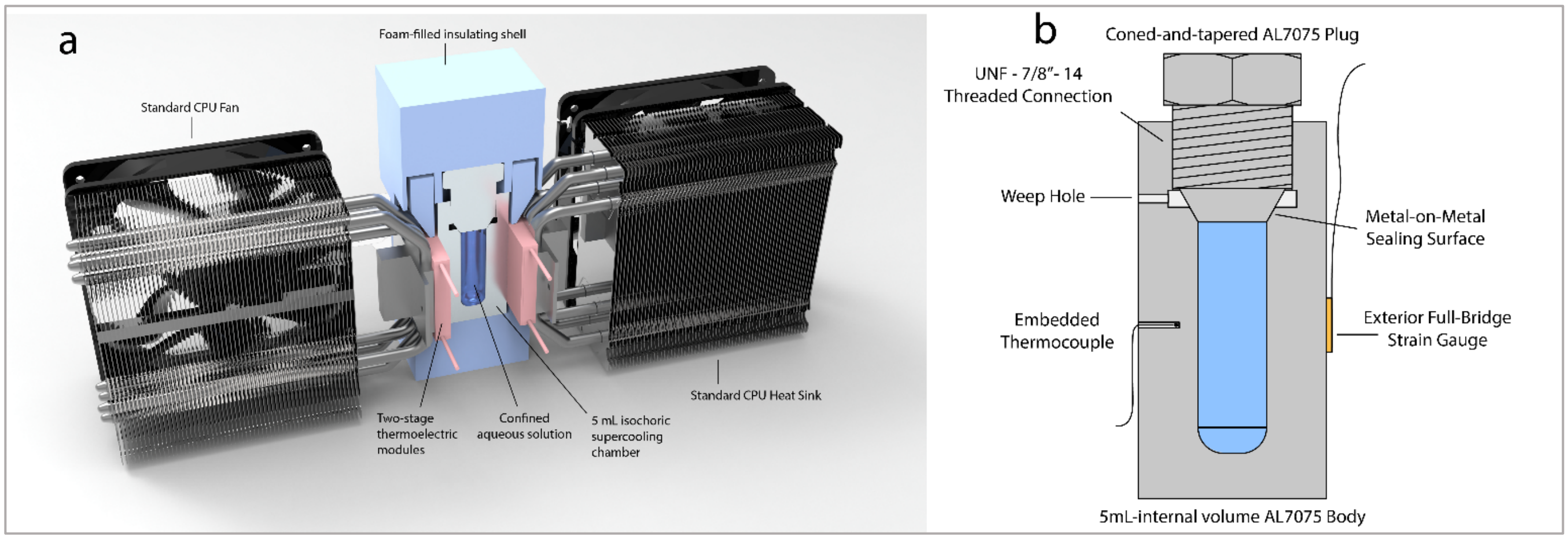

**Figure 1: Schematic of the Isochoric Nucleation Detector (INDe).** (a) 3D rendering depicting cutaway of INDe chamber, insulation shell and temperature control assemblies consisting of thermoelectric modules and fan-cooled CPU heat sinks. (b) 2D cutaway schematic of 5mL INDe isochoric chamber depicting sealing mechanism, embedded thermocouple and strain gauge.

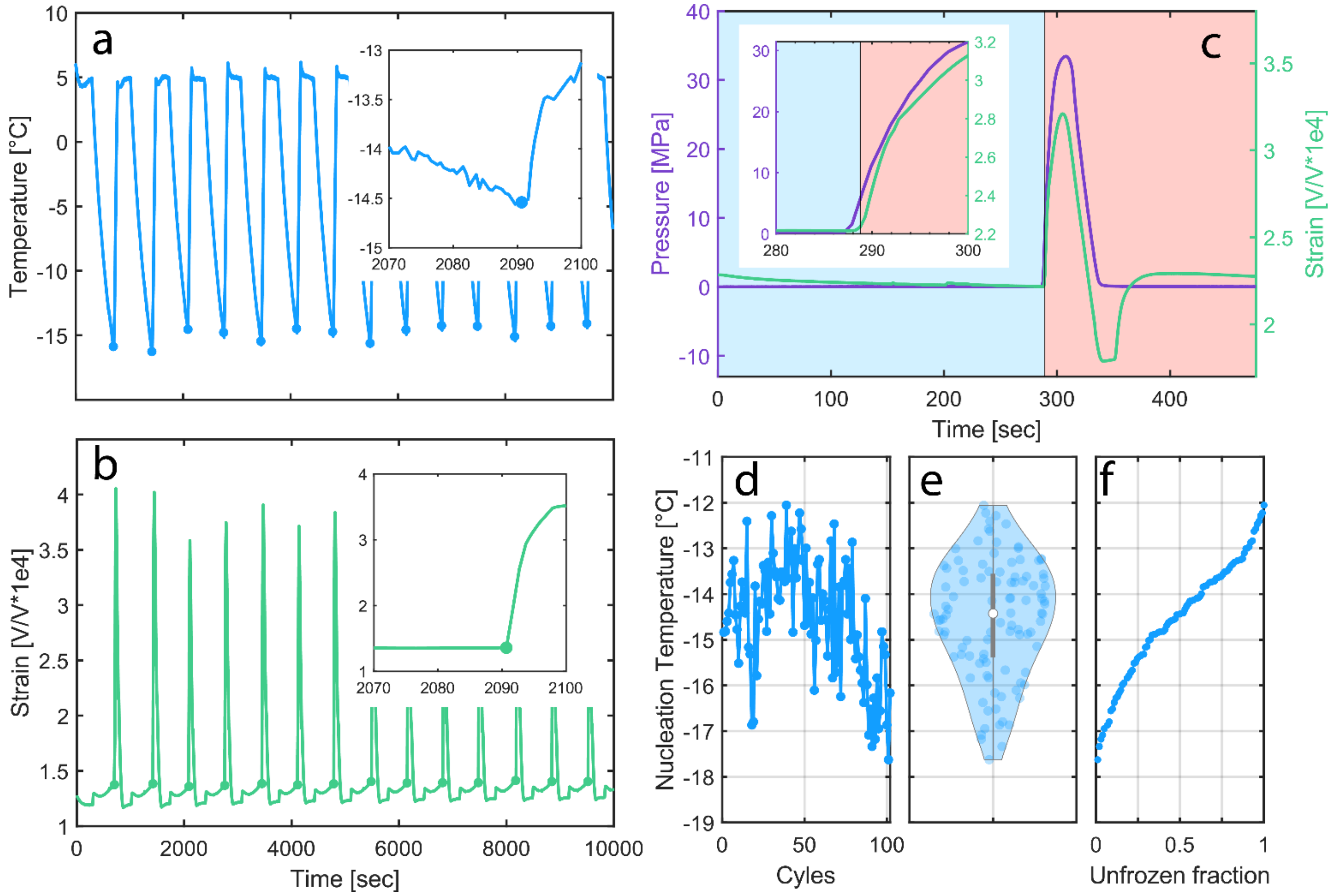

**Figure 2: Example data obtained by the INDe**. (a) Raw temperature curves for a series of cooling/warming cycles. Markers at bottom of saw-tooth indicate nucleation temperature. Inset depicts zoom-in on one nucleation event. (b) Corresponding raw strain curve. Spike in signal caused by nucleation of ice within isochoric chamber. Markers at base of spike indicate nucleation event. Inset depicts zoom-in on one nucleation event. (c) Validation of equivalence between pressure and strain monitoring for nucleation detection. Blue region indicates cooling period. Red region indicates warming period after detection of nucleation. (d) Representative extracted nucleation temperatures from one INDe experiment. (e) Violin plot representation of nucleation temperature distribution. (f) Survivor curve representation of nucleation temperature distribution.

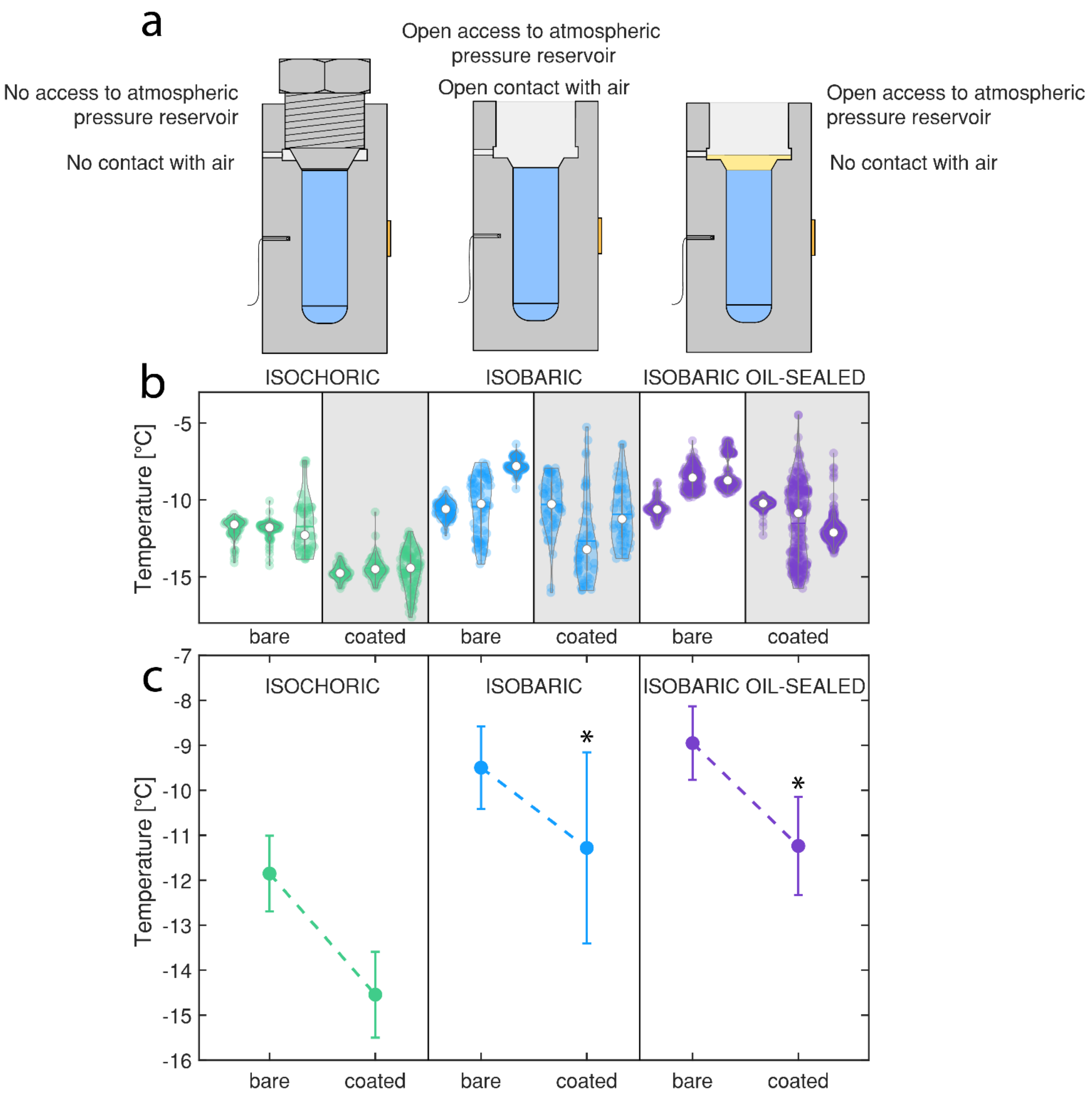

**Figure 3: Investigation of thermodynamic conditions (isochoric, conventional isobaric, and isobaric oil-sealed) with two different wall conditions (bare metal and petrolatum-coated).** (a) Schematic illustration of experimental configurations. (b) Violin plot distributions of nucleation temperatures. (c) Mean nucleation temperature for each experimental condition. For each condition, experiments were conducted in three (3) separate chambers for a minimum of 50 cycles each. Each value reported in (b) is the average of the mean nucleation temperatures from each chamber, weighted by the number of cycles. Error bars indicate the average of the standard deviations for each chamber, weighted by the number of cycles. Asterisks (*) indicate conditions that produced statistically similar results.

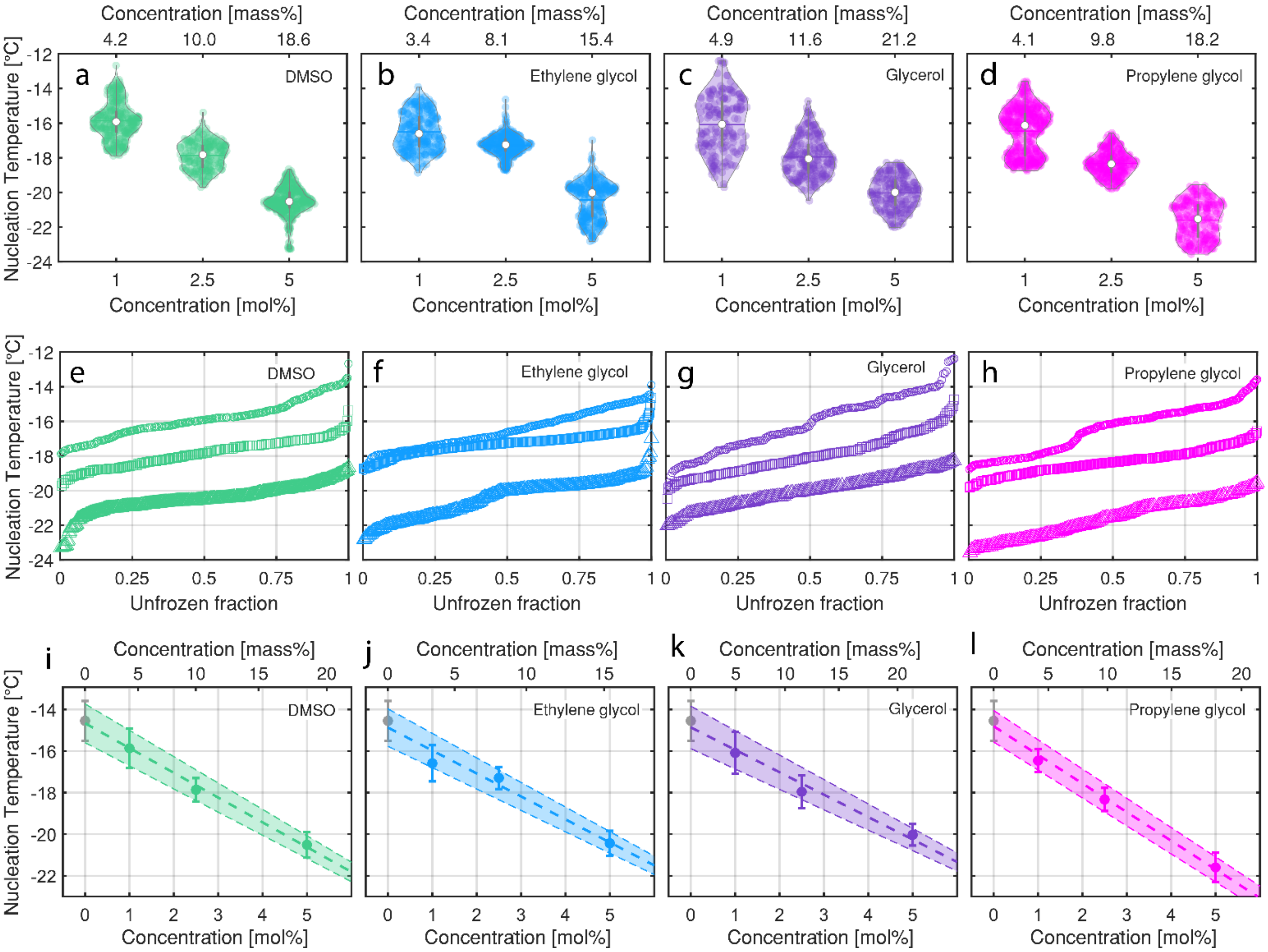

**Figure 4: Nucleation temperature data for binary solutions of water and four solutes: DMSO (green), ethylene glycol (blue), glycerol (purple), propylene glycol (pink).** (a)-(d) Violin plot distributions of nucleation temperatures. (e)-(h) Survivor curves for distributions shown in (a)-(d). (i)-(l) Weighted mean nucleation temperatures as a function of concentration. Error bars and shaded region indicate one standard deviation.

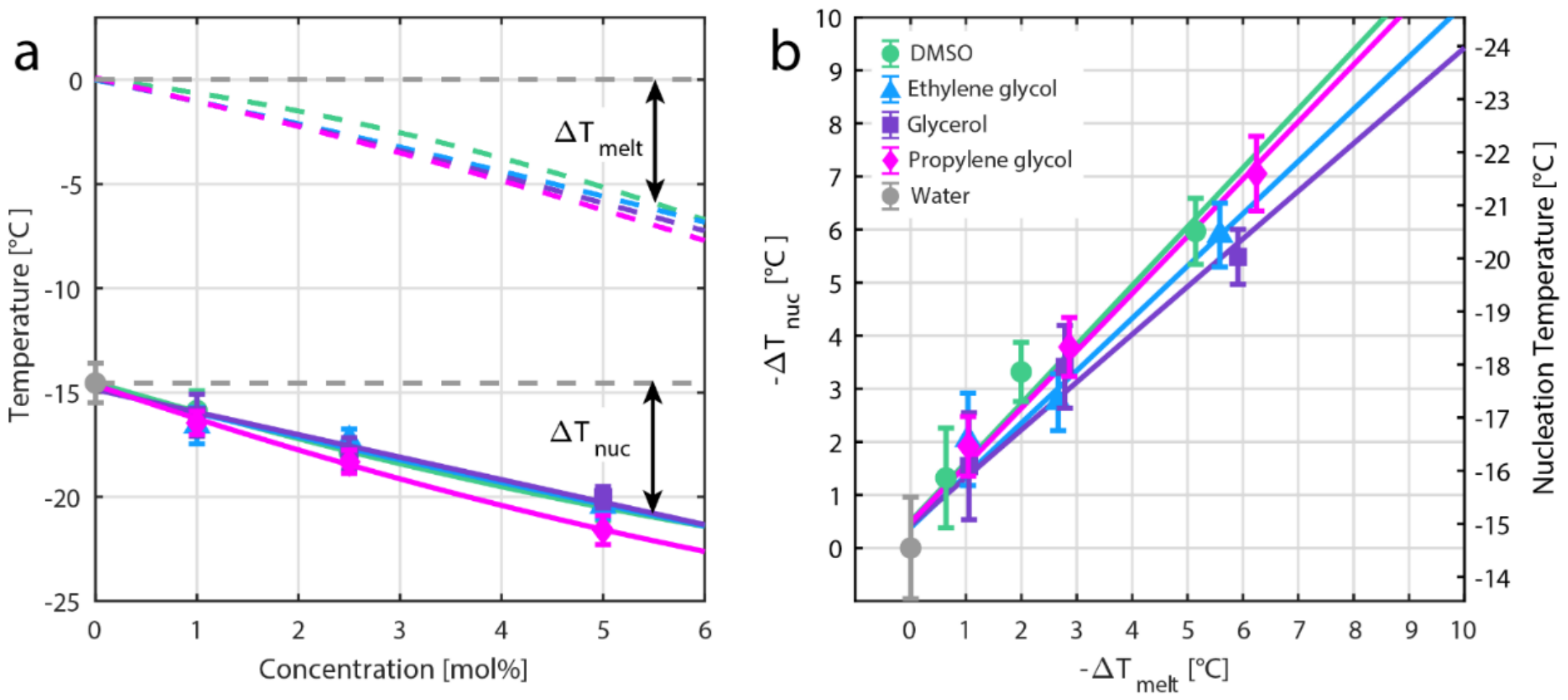

**Figure 5: Equilibrium melting temperature and nucleation temperature curves for four solutes: DMSO (green), ethylene glycol (blue), glycerol (purple), and propylene glycol (pink).** (a) Equilibrium melting temperatures and nucleation temperatures as a function of concentration. $\Delta T_{melt}$ is the equilibrium melting point depression. $\Delta T_{nuc}$ is the difference between the nucleation temperature for the solutions and the nucleation temperature of pure water. (b) Change in nucleation temperature versus melting point depression.

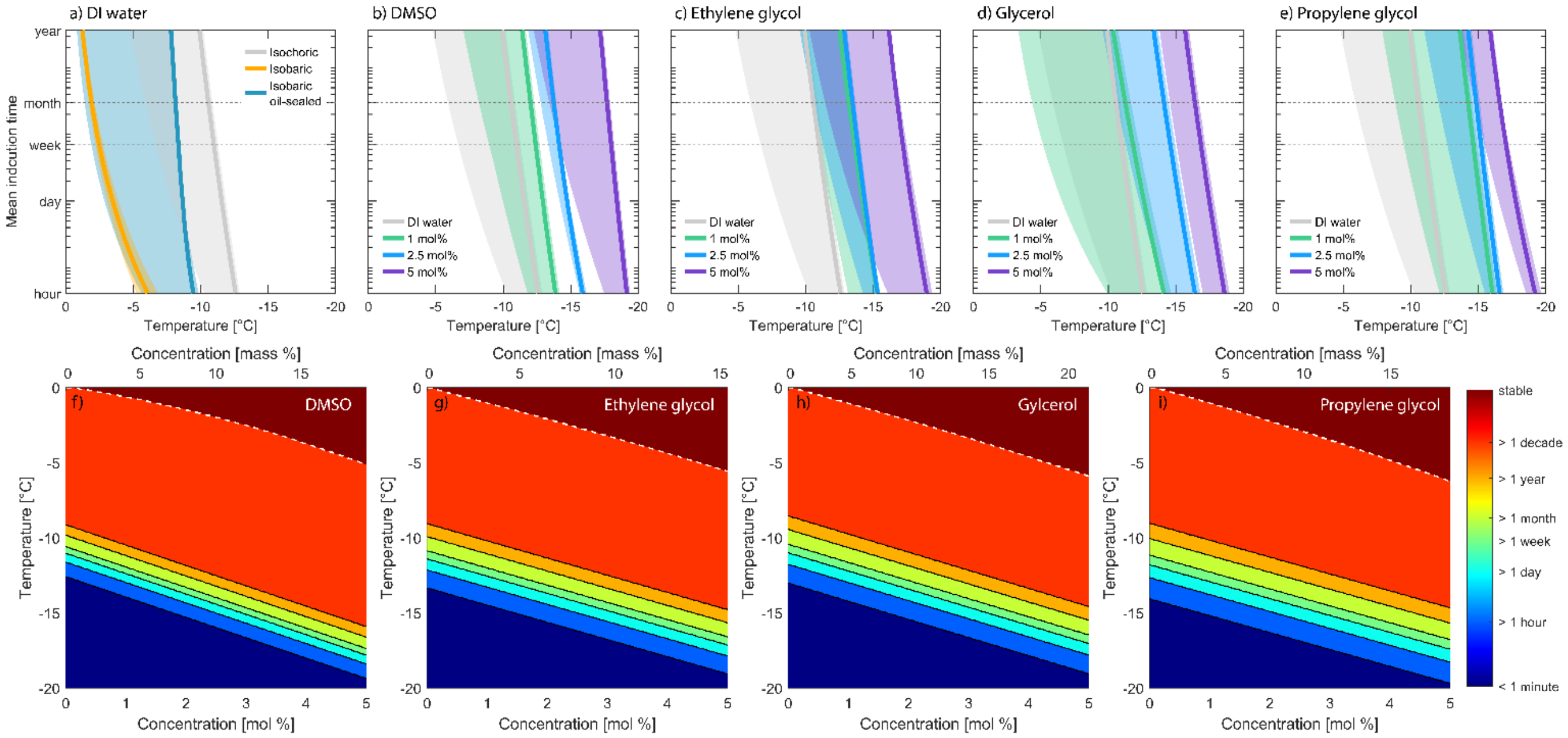

**Figure 6: Nucleation induction times.** (a) Nucleation induction times as a function of temperature for pure water under three thermodynamic boundary conditions (isochoric, isobaric, isobaric oil-sealed) and with petrolatum-coated walls. (b)-(e) Nucleation induction times for solutions of DMSO (green), ethylene glycol (blue), glycerol (purple) and propylene glycol (pink) at concentrations of 1 mol%, 2.5 mol% and 5 mol%. Shaded region represents range of induction times from the three individual trials for each condition. Solid lines represent the average of the computed induction times. (f)-(i) Induction time stability maps as a function of temperature and solution concentration. White dashed lines indicate equilibrium melting point, above which the solutions are indefinitely stable.

## Supplemental Information for:

# Methods to stabilize aqueous supercooling identified by use of an isochoric nucleation detection (INDe) device

Anthony Consiglio, Drew Lilley, Boris Rubinsky, Matthew J. Powell-Palm†

*Department of Mechanical Engineering, University of California at Berkeley, Berkeley, CA 94720*

†*To whom correspondence should be addressed:* mpowellp@berkeley.edu

## 1 Isochoric nucleation detection (INDe) array

The isochoric nucleation detection (INDe) device is a scalable platform suited for performing high-throughput supercooling experiments on aqueous systems. The INDe system introduced in this study utilizes solid-state thermoelectric modules for both cooling and heating. Together with air-cooled heat sinks, this design enables implementation in any laboratory setting without the need for refrigerated fluid circulating systems. Pictured in Figure S1 is an array of six (6) individual INDe devices operating off a single Raspberry Pi computer (bottom right). The temperature of each INDe is controlled by a separate PID temperature controller (bottom left), which receives commands from the centralized control dashboard.

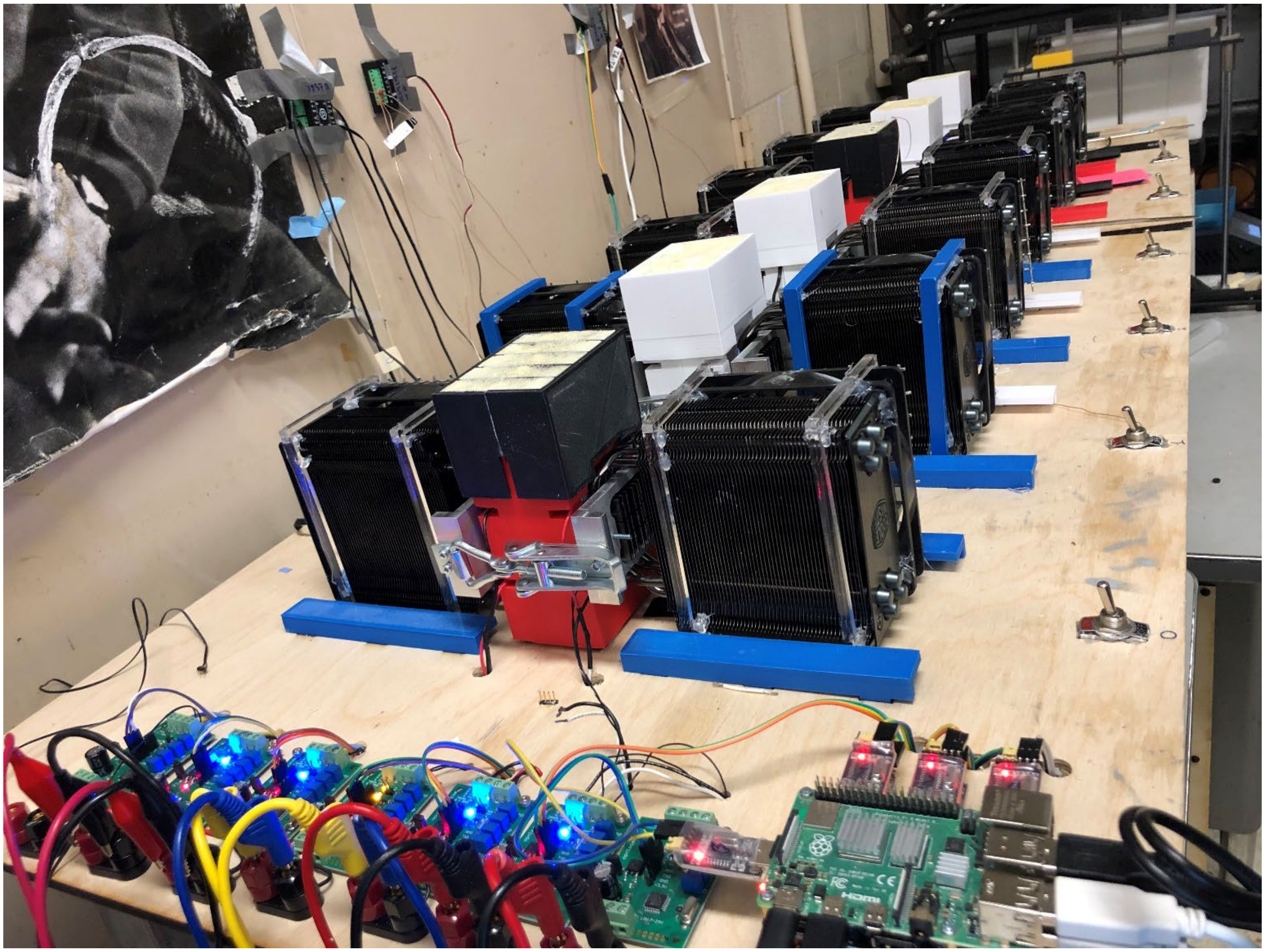

*Figure S1: INDe array*

## 2 Solution experiment raw data violin plots

Shown in Figure S2 are the violin plot distributions of the nucleation temperatures for the 36 individual solution experiments. Experiments were conducted in three independent chambers for each of the four solutes (DMSO, ethylene glycol, glycerol and propylene glycol) at three concentrations (1 mol%, 2.5 mol% and 5 mol%).

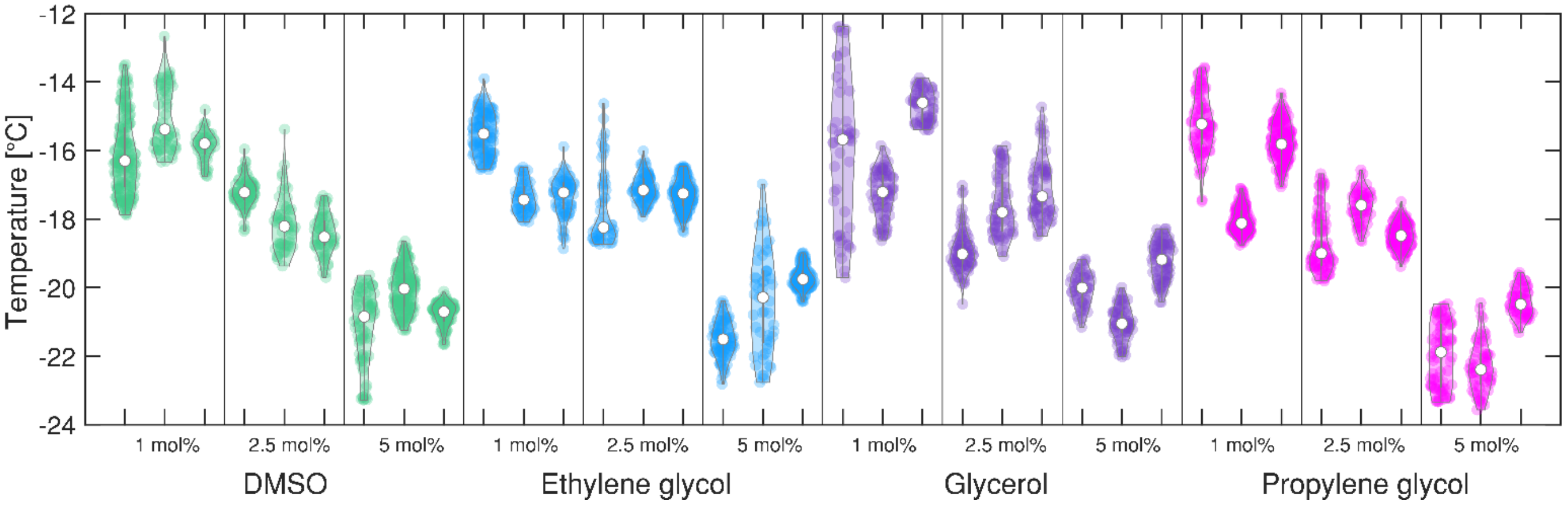

*Figure S2: Nucleation temperature violin plot distributions for solution experiments*

## 3 Statistical model for calculation of isothermal induction time

As described by Lilley et al.[1], transient supercooling experiments of the type performed in this study, wherein the temperature of a sample is lowered at constant rate until nucleation occurs, may be modelled as a non-homogeneous Poisson process. For surface-initiated nucleation, the survivor function, $\chi(T)$, is related to the nucleation rate, $J(T)$, by Equation 3 of the main text as

$$\chi(T) = e^{-\frac{1}{\beta}\int_{T_m}^{T} J(T)dT} = e^{-\frac{1}{\beta}\int_{T_m}^{T} \gamma(T_m-T)^n dT} = e^{-\frac{1}{\beta}\frac{\gamma(T_m-T)^{1+n}}{1+n}}$$

wherein $\beta$ is the cooling rate, $T_m$ is the equilibrium melting point, and $\gamma$ and $n$ are empirical fitting parameters. This equation may be fitted to the survivor curves generated by the INDe in order to obtain values for $\gamma$ and $n$. Figure S3 shows the experimental (markers) and fitted (solid lines) survivor curves for each of the trials. Table S1 provides the values for the fitting parameters, $\gamma$ and $n$.

Supplemental Information

*Table S1: Empirical nucleation rate parameters obtained by fitting experimental survivor function data to Poisson distribution*

| | $n$ | $\gamma$ |
|---|---|---|
| Isochoric DI water Bare | 30.25 | 1.67E-34 |
| | 54.52 | 1.69E-60 |
| | 7.8 | 6.68E-11 |
| Isochoric DI water Petrolatum | 38.64 | 3.37E-47 |
| | 27.95 | 1.32E-34 |
| | 11.87 | 2.68E-16 |
| Isobaric DI water Bare | 21.86 | 1.35E-24 |
| | 5.56 | 2.10E-08 |
| | 20.79 | 2.10E-20 |
| Isobaric DI water Petrolatum | 5.88 | 1.69E-08 |
| | 5.67 | 5.29E-09 |
| | 5.67 | 1.69E-08 |
| Isobaric oil-sealed DI water Bare | 26.4 | 4.12E-29 |
| | 12.88 | 2.64E-14 |
| | 8.34 | 6.68E-10 |
| Isobaric oil-sealed DI water Petrolatum | 59.33 | 1.03E-61 |
| | 4.7 | 8.33E-08 |
| | 24.11 | 3.29E-28 |

| | $n$ | $\gamma$ |
|---|---|---|
| DMSO 1 mol% | 15.13 | 2.10E-20 |
| | 17.74 | 6.68E-23 |
| | 44.20 | 2.64E-54 |
| DMSO 2.5 mol% | 49.17 | 4.21E-60 |
| | 27.69 | 1.06E-35 |
| | 33.46 | 8.42E-43 |
| DMSO 5 mol% | 16.52 | 2.68E-22 |
| | 26.24 | 5.29E-33 |
| | 60.40 | 8.42E-74 |
| Ethylene glycol 1 mol% | 26.62 | 4.21E-33 |
| | 42.76 | 6.68E-54 |
| | 37.25 | 2.64E-47 |
| Ethylene glycol 2.5 mol% | 17.42 | 8.33E-23 |
| | 43.99 | 5.29E-53 |
| | 37.20 | 2.13E-45 |
| Ethylene glycol 5 mol% | 34.85 | 5.29E-44 |
| | 11.28 | 8.42E-16 |
| | 52.65 | 2.10E-62 |

| | $n$ | $\gamma$ |
|---|---|---|
| Glycerol 1 mol% | 6.68 | 1.34E-10 |
| | 26.56 | 2.13E-34 |
| | 32.71 | 3.37E-39 |
| Glycerol 2.5 mol% | 35.92 | 1.69E-45 |
| | 17.53 | 8.33E-23 |
| | 19.29 | 1.34E-24 |
| Glycerol 5 mol% | 32.12 | 4.21E-39 |
| | 34.85 | 3.33E-43 |
| | 26.72 | 3.37E-32 |
| Propylene glycol 1 mol% | 18.76 | 6.68E-24 |
| | 51.63 | 1.67E-65 |
| | 28.65 | 1.34E-35 |
| Propylene glycol 2.5 mol% | 20.63 | 5.29E-27 |
| | 37.57 | 5.29E-46 |
| | 49.23 | 1.05E-60 |
| Propylene glycol 5 mol% | 17.80 | 1.06E-23 |
| | 25.92 | 1.67E-33 |
| | 39.23 | 3.33E-47 |

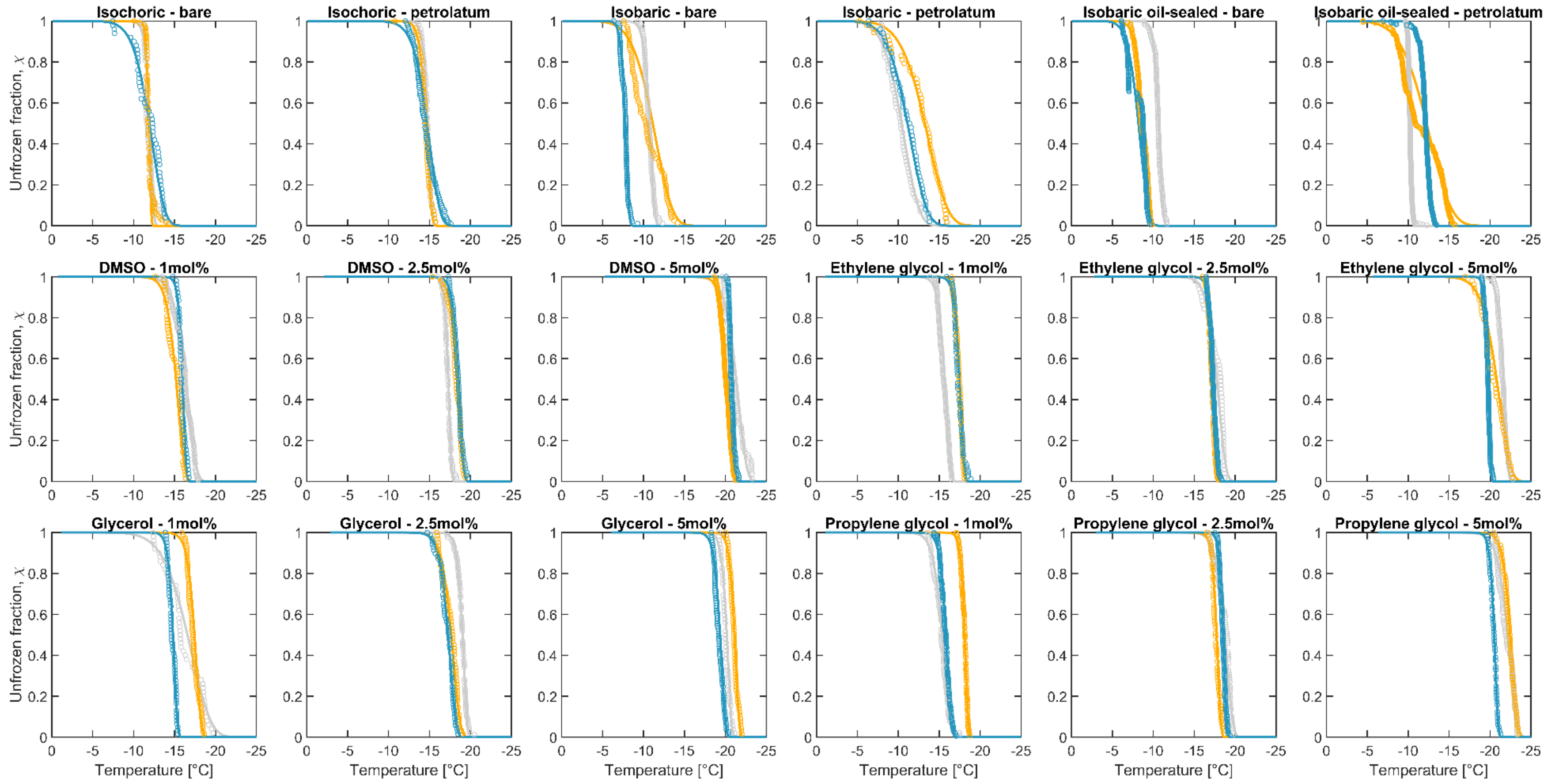

*Figure S3: Experimental (markers) and fitted (solid lines) survivor curves.*